\documentclass[a4paper]{article}
\begin{document}

\title{\LARGE \bf Unfinished revolution\\[2mm]
\rm \normalsize\em Introductive chapter of a book on Quantum Gravity, 
edited by Daniele Oriti,\\ to appear with
Cambridge University Press}

\author{Carlo Rovelli\\
{\small \em Centre de Physique Th\'eorique de Luminy%
\footnote{Unit\'e mixte de recherche (UMR 6207) du CNRS et des Universit\'es
de Provence (Aix-Marseille I), de la Meditarran\'ee (Aix-Marseille II) et du Sud (Toulon-Var); laboratoire affili\'e \`a la FRUMAM (FR 2291).}, case 907, F-13288 Marseille, EU}}
\date{\small\today}
\maketitle

One hundred and forty-four years elapsed between the publication of Copernicus's \emph{De Revolutionibus}, which opened the great scientific revolution of the XVII century, and the 
publication of Newton's \emph{Principia}, the final synthesis that brought that revolution
to a spectacularly successful end.  During those hundred and forty-four years,
the basic grammar for understanding the physical world changed and the 
old picture of reality was reshaped in depth. 

At the beginning of the XX century,  General Relativity (GR) and 
Quantum Mechanics (QM) once again began reshaping our 
basic understanding of space and time and, respectively, matter, 
energy and causality ---arguably to a no lesser extent.  But we have 
not been able to combine these new insights into a novel coherent 
synthesis, yet.   The XX century scientific revolution opened by 
GR and QM is therefore still wide open.  
We are in the middle of an unfinished scientific revolution.  Quantum 
Gravity is the tentative name we give to the ``synthesis to be found". 

In fact, our present understanding of the physical world at the fundamental level is in a state 
of great confusion.  The present knowledge of the elementary dynamical laws 
of physics is given by the application of QM to fields, namely quantum
field theory (QFT), by the particle--physics Standard Model (SM), and by GR. 
This set of fundamental theories has obtained an empirical
success nearly unique in the history of science: so far there isn't any clear 
evidence of observed phenomena that clearly escape or contradict
this set of theories ---or a minor modification of the same, such as a 
neutrino mass or a cosmological constant.%
\footnote{Dark matter (not dark energy) might perhaps be contrary evidence.}
But, the theories in this set are based on badly self-contradictory 
assumptions. In GR the gravitational field is assumed to be a 
classical deterministic dynamical field, identified with 
the (pseudo) riemannian metric of spacetime: but with QM we have
understood that all 
dynamical fields have quantum properties.  The other way around,
conventional QFT relies heavily on global Poincar\'e invariance and on 
the existence of a non--dynamical background spacetime metric: 
but with GR we have understood that there is no such non--dynamical 
background spacetime metric in nature. 

In spite of their  empirical success, GR and QM offer 
a schizophrenic and confused understanding of the physical world.
The conceptual foundations of classical GR are contradicted 
by QM and the conceptual foundation of conventional QFT are 
contradicted by GR. Fundamental physics is today 
in a peculiar phase of deep conceptual confusion.  

Some deny that such a major internal contradiction in our picture 
of nature exists.   On the one hand, some refuse to take QM seriously.  
They insist that QM makes no 
sense, after all, and therefore the fundamental world must be 
essentially classical. This doesn't put us in a better shape,
as far as our understanding of the world is concerned. 

Others, on the other hand, and in particular some
hard--core particle physicists, do not accept the lesson of GR. They 
read GR as a field theory that can be consistently formulated in full 
on a fixed metric background, and treated  within conventional QFT 
methods. They motivate this refusal  by insisting than GR's insight 
should not be taken too seriously, because GR is just a low--energy limit 
of a more fundamental theory.  In doing so, they confuse the details of
the Einstein's equations (which might well be modified at high 
energy), with the new understanding of space and time brought
by GR. This is coded in the background independence of the 
fundamental theory and expresses Einstein's discovery that 
spacetime is not a fixed background, as it was assumed in 
special relativistic physics, but rather a dynamical 
field.

Nowadays this fact is finally being recognized even by those who 
have long refused to admit that GR forces a revolution in the way to 
think about space and time, such as some of the leading voices in
string theory.  In a recent interview \cite{gross}, for instance, Nobel 
laureate David Gross says:
`` [...] this revolution will likely change the way we think about space 
and time, maybe even eliminate them completely as a basis for our 
description of reality". This is of course something that has been 
known since the
 1930's \cite{bronstein} by  anybody who has
taken seriously the problem of the implications of GR and QM.  The 
problem of the conceptual novelty of GR, which the string approach 
has tried to throw out of the door, comes back by the window.

The scientists trying to resist quantum theory or background
independence remind me of Tycho Brahe, who tried hard
to conciliate Copernicus advances with the ``irrefutable evidence" that 
the Earth is immovable at the center of the universe.  To let the 
background spacetime go is perhaps as difficult as
letting go of the unmovable background Earth. 
The world may not be the way it appears 
in the tiny garden of our daily experience.  

Today, many scientists do not hesitate to take seriously speculations such as extra 
dimensions, new symmetries or multiple universes, for which there isn't a wit
of empirical evidence; but refuse to take seriously the conceptual implications
of the physics of the XX century with the enormous body of empirical evidence
supporting them.   Extra dimensions, new symmetries,  
multiple universes and the like, still make perfectly sense in a pre-GR, pre-QM,  
Newtonian world, while to take GR and QM seriously \emph{together} requires a genuine 
reshaping of our world view. 

After a century of empirical successes that have only equals in Newton's and 
Maxwell's theories, it is time to take seriously 
GR and QM, with their full conceptual implications; to find a
way of thinking the world in which what we have learned with QM 
and what we have learned with GR make sense together  ---finally
bringing the XX century scientific revolution to its end. This is the 
problem of Quantum Gravity. 

\section{Quantum spacetime}

Roughly speaking, we learn from GR that spacetime is a dynamical
field and we learn from QM that all dynamical field are quantized. 
A quantum field has a granular structure, and a probabilistic dynamics, 
that allows quantum superposition of different states. 
Therefore at small scales we might expect a ``quantum spacetime"
formed by ``quanta of space" evolving probabilistically, and 
allowing ``quantum superposition of spaces".  The problem of 
quantum gravity is to give a precise mathematical and physical meaning to this 
vague notion of  ``quantum spacetime".

Some general indications about the nature of quantum spacetime, and on
the problems this notion raises, can be obtained from elementary
considerations. The size of quantum mechanical
effects is determined by Planck's constant $\hbar$.  The strength of
the gravitational force is determined by Newton's constant $G$, and
the relativistic domain is determined by the speed of light $c$.  By
combining these three fundamental constants we obtain the 
Planck length $l_{\rm P}=\sqrt{\hbar G/c^3}\sim 10^{-33}$ cm. 
Quantum-gravitational effects are likely to be negligible at distances
much larger than $l_{\rm P}$, because at these scales we can neglect
quantities of the order of $G$, $\hbar$ or $1/c$.  

Therefore we expect
the classical GR description of spacetime as a pseudo-riemannian 
space to hold at 
scales larger than $l_{\rm P}$, but to break down approaching this scale,
where the full structure of quantum spacetime becomes relevant.  
Quantum gravity is
therefore the study of the structure of spacetime at the Planck scale.

\subsection{Space}

Many simple arguments indicate that $l_{\rm P}$ may play the role of a
\emph{minimal} length, in the same sense in which $c$ is the maximal
velocity and $\hbar$ the minimal exchanged action.  

For instance,
the Heisenberg principle requires that the position of an object of
mass $m$ can only be determined with uncertainty $x$ satisfying
$mvx>\hbar$, where $v$ is the uncertainty in the velocity; special
relativity requires $v<c$; and according to GR there is a limit to
the amount of mass we can concentrate in a region of size $x$, given
by $x>Gm/c^2$, after which the region itself collapses into a black
hole, subtracting itself from our observation.  Combining these
inequalities we obtain $x>l_{P}$. That is, gravity, relativity and
quantum theory, taken together, appear to prevent position to be
determined more precisely than the Planck scale.  

A number of
considerations of this kind have suggested that space might not be
infinitely divisible.  It may have a quantum granularity at the
Planck scale, analogous to the granularity of the energy in a
quantum oscillator.  This granularity of space is fully realized in
certain quantum gravity theories, such as loop quantum gravity, and there are
hints of it also in string theory. Since this is a quantum granularity, 
it escapes the traditional objections to the atomic nature of space. 

\subsection{Time}

Time is affected even more radically by the quantization of gravity. 
In conventional QM, time is treated as an external parameter and
transition probabilities change in time.  In GR there is no external
time parameter.  Coordinate time is a gauge variable which is not
observable, and the physical variable measured by a clock is a
nontrivial function of the gravitational field.  Fundamental
equations of quantum gravity might therefore not be written as evolution equations
in an observable time variable.  And in fact, in the quantum--gravity
equation  \emph{par excellence}, the Wheeler-deWitt
equation, there is no time variable $t$ at all. 

Much has been written on the fact that the equations of
nonperturbative quantum gravity  do not contain the time variable $t$.  This 
presentation of the ``problem of time in quantum gravity", however, is a bit 
misleading, since it mixes a problem of classical GR with a specific
quantum gravity issue. Indeed, \emph{classical} GR as well
can be entirely formulated in the Hamilton-Jacobi formalism, where no 
time variable appears either.

In classical GR, indeed, the notion of time
differs strongly from the one used in the special-relativistic context.  
Before special relativity, one assumed that there is a universal 
physical variable $t$, measured by clocks, such that all physical 
phenomena can be described in terms of evolution equations in 
the independent variable $t$.  
In \emph{special} relativity, this notion of time is weakened.  Clocks
do not measure a universal time variable, but only the proper time
elapsed along inertial trajectories.  If we fix a Lorentz frame, nevertheless, 
we can still describe all physical phenomena in terms of
evolution equations in the independent variable $x^0$, even though
this description hides the covariance of the system.

In \emph{general} relativity, when we describe the dynamics of the 
gravitational field (not to be confused with the dynamics of matter in
a given gravitational field), there is \emph{no} external time variable 
that can play the role of observable independent evolution variable.  
The field equations are written in terms of an evolution parameter, 
which is the time coordinate $x^0$, but this coordinate, does not 
correspond to anything directly observable. The proper time $\tau$ 
along spacetime trajectories cannot be used as an independent variable 
either, as $\tau$ is a complicated non-local function of the gravitational 
field itself.  Therefore, properly speaking, GR does not admit a description 
as a system evolving in terms of an observable time variable. 
This does not mean that GR lacks predictivity.  Simply put, what
GR predicts are relations between (partial) observables, which in
general cannot be represented as the evolution of dependent 
variables on a preferred independent time variable.

This weakening of the notion of time in classical GR is rarely
emphasized: After all, in classical GR we may disregard
the full dynamical structure of the theory and consider only
individual solutions of its equations of motion.  A single
solution of the GR equations of motion determines ``a spacetime", 
where a
notion of proper time is associated to each timelike worldline.  

But in the quantum context a single solution of the dynamical equation is 
like a single ``trajectory" of a quantum particle: in quantum theory there are no 
physical individual 
trajectories: there are only transition probabilities between
observable eigenvalues.  Therefore in quantum gravity it is likely to be 
impossible to describe the world in terms of a spacetime, in 
the same sense in which the motion of a quantum electron 
cannot be described in terms of a single trajectory. 

To make sense of the world at the Planck scale, and to find a consistent 
conceptual framework for GR and QM, we might have to give up the 
notion of time altogether, and
learn ways to describe the world in atemporal terms.  Time might be a
useful concept only within an approximate description of the physical
reality.

\subsection{Conceptual issues}

The key difficulty of quantum gravity
may therefore be to find a way to understand the physical world in the
absence of the familiar stage of space and time.  
What might be needed is to free ourselves from the 
prejudices associated with the habit of
thinking of the world as ``inhabiting space" and ``evolving in time".

Technically, this means that the quantum states of the gravitational
field cannot be interpreted like the $n$-particle states of conventional
QFT as living on a given spacetime.  Rather, these quantum states 
must themselves determine and define a spacetime ---in the 
manner in which the classical solutions of GR do. 

Conceptually, the key question is 
whether or not it is logically possible to understand the world in the
absence of fundamental notions of time and time evolution, and whether or not 
this is consistent with our experience of the world.   

The difficulties  of quantum gravity are indeed largely conceptual. 
Progress  in quantum gravity cannot be just technical.  The search for a quantum
theory of gravity raises once more old questions such as:  What is space? 
What is time?   What is the meaning of ``moving''?  Is motion to be defined with respect to
objects or with respect to space?  And also: What is causality? 
What is the role of the observer in physics?
Questions of this kind have played a central role in periods of major
advances in physics.  For instance, they played a central role for
Einstein, Heisenberg, and Bohr.  But also for
Descartes, Galileo, Newton and their contemporaries, as well as for Faraday and
Maxwell.  

Today some physicists view this manner
of posing problems as ``too philosophical".  Many physicists of the
second half of the twentieth century, indeed, have viewed questions of
this nature as irrelevant.  This view was appropriate for the problems
they were facing.  When the basics are clear and the issue is
problem-solving within a given conceptual scheme, there is no reason
to worry about foundations: a pragmatic approach is the most effective
one.  Today the kind of difficulties that fundamental physics faces
have changed.  To understand quantum spacetime, physics has to return,
once more, to those foundational questions.  

\section{Where are we?}

Research in quantum gravity has developed slowly for several decades during the
XX century, because GR had little impact on
the rest of physics and the interest of many theoreticians was
concentrated on the development of quantum theory and particle
physics.  In the last twenty years, the explosion of empirical
confirmations and concrete astrophysical, cosmological and even
technological applications of GR on the one hand,
and the satisfactory solution of most of the particle physics
puzzles in the context of the SM on
the other, have led to a strong concentration of interest in quantum
gravity, and the progress has become rapid. Quantum gravity is
viewed today by many as the big open challenge in fundamental physics. 

Still, after 70 years of research in quantum gravity, there is no consensus, 
and no established theory. I think it is fair to say that there isn't even a single
complete and consistent \emph{candidate} for a quantum theory of gravity.

In the course of 70 years, numerous 
ideas have been explored, fashions have come and gone, the discovery 
of the Holy Grail of quantum gravity has been several times announced, only to be later 
greeted with much scorn. Of the tentative theories studied today (strings, 
loops and spinfoams, non-commutative geometry, dynamical triangulations 
or other), each is to a large extent incomplete and 
none has yet received a whit  of direct or indirect empirical  
support.

However, research in quantum gravity has not been meandering
meaninglessly.  On the contrary, a consistent logic has guided the
development of the research, from the early formulation of the problem
and of the major research programs in the fifties to nowadays.  The
implementation of these programs has been laborious, but has been
achieved.  Difficulties have appeared, and solutions have been
proposed, which, after much difficulty, have lead to the realization,
at least partial, of the initial hopes.  

It was suggested in the early seventies that GR could perhaps be
seen as the low energy limit of a Poincar\'e invariant QFT without uncontrollable
divergences  \cite{zumino}; and today, 30 years later, a theory likely to have these
properties ---perturbative string
theory--- is known.  It was also suggested  in the early seventies that
non-renormalizability might not be fatal for quantum GR \cite{parisi,weinberg76}
and that the Planck scale could
cut divergences off nonperturbatively by inducing a quantum discrete structure
of space; and today we know that this is in fact the case ---ultraviolet finiteness is 
realized precisely in this manner in canonical loop quantum gravity and in 
some spinfoam models.    In 1957 Charles Misner indicated that in the
canonical framework one should be able to compute quantum 
eigenvalues of geometrical quantities \cite{misner}; and
in 1995, 37 years later, eigenvalues of area and volume were computed 
---within loop quantum gravity \cite{eigen}.  Much remains to be understood and some of the
current developments might lead nowhere. But looking at
the entire development of the subject, it is difficult to deny that
there has been substantial progress. 

In fact, at least two major research programs can today claim to have, if not
a complete candidate theory of quantum gravity, at least a large piece of it: 
string theory (in its perturbative and still incomplete nonperturbative versions) and loop quantum gravity (in its canonical as well as covariant
 --spinfoam-- versions) are both incomplete theories, full of defects ---in general,  
 strongly emphasized within the opposite camp--- and 
without any empirical support, but they are both remarkably rich and coherent 
theoretical frameworks, that \emph{might} not be far from the solution of the puzzle. 

Within these frameworks, classical and long 
intractable, physical, astrophysical and cosmological quantum gravity problems
can finally be concretely treated. Among these: black hole's entropy 
and fate, the physics of the big--bang singularity and the way it has 
affected the currently observable universe, and many others. 
Tentative predictions are being developed, and the attention to the 
concrete possibility of testing these predictions with observations that 
could probe the Planck scale is very alive. All this was unthinkable only 
a few years ago. 

The two approaches differ profoundly in their hypotheses,
achievements, specific results, and in the conceptual frame they
propose.  The issues they raise concern the foundations of the
physical picture of the world, and the debate between the two
approaches involves conceptual, methodological and philosophical
issues.

In addition, a number of other ideas, possibly alternative, possibly 
complementary to the two best developed theories and to one another, 
are being explored.  These include noncommutative geometry,
dynamical triangulation, effective theories, causal sets and many others. 

The possibility 
that \emph{none} the currently explored hypotheses will eventually turn out to 
be viable, or, simply, none will turn out to be the way chosen by Nature, is very 
concrete, and should be clearly kept in mind.   But the rapid and multi--front
progress of the last years raises hopes.  Major well--posed open questions 
in theoretical physics (Copernicus or Ptolemy? Galileo's parabolas or Kepler's 
ellipses? How to describe electricity and magnetism?  Does Maxwell theory 
pick a preferred reference frame? How to do the quantum mechanics of 
interacting fields...?) have rarely been solved in a few years.  But they have
rarely resisted more than a few decades.   Quantum gravity ---the problem of
describing the quantum properties of spacetime-- is one of these major
problems, and it is reasonably well defined: is there a coherent theoretical 
framework consistent with quantum theory and with general relativity?  It
is a problem which has been on the table since the thirties, but it is only in the last
couple of decades that the efforts of the theoretical physics community have 
concentrated on it.    

Maybe the solution is not far. In any case, we 
are not at the end of the road of physics, we are half-way through the woods
along  a major scientific revolution.

Thanks to Randy Reeder for the editing.

\subsubsection*{Bibliographical note}

For details on the history of quantum gravity see the historical appendix in
\cite{book}; and, for early history see \cite{stachel} and
\cite{gore}.  For orientation on current research on quantum gravity, see the
review papers \cite{horowitz,carliprev,isham,reviewqg}.   As a
general introduction to quantum gravity ideas, see the old classic reviews, which
are rich in ideas and present different points of view, such as John
Wheeler 1967 \cite{wheeler67}, Steven Weinberg 1979 \cite{weinberg76},
Stephen Hawking 1979 and 1980 \cite{haw,haw2}, Karel Kuchar 1980
\cite{Kuchar}, and Chris Isham's magisterial syntheses
\cite{chris,Isham80,Isham96}. On string theory, classic textbooks are
Green, Schwarz and Witten, and Polchinksi \cite{Strings}.  On loop quantum gravity,
including the spinfoam formalism, see \cite{book,thomas,ashtekar}, or the
older papers \cite{lqg}. On spinfoams see also
\cite{Perez}.  On noncommutative geometry \cite{conneslibro} and on 
dynamical triangulations \cite{Carfora}.  For a discussion of the 
difficulties of string theory and a comparison of the results of strings and 
loops, see \cite{dialog}, written in the form of a dialogue, and
\cite{leereview}.  On the more philosophical challenges raised by
quantum gravity, see \cite{meets}.
Smolin's popular book \cite{LeeBook} provides a
readable introduction to quantum gravity. 
The expression ``half way through the woods" to characterize the present state
of fundamental theoretical physics is taken from  \cite{halfway}. 
My own view on quantum gravity is developed in detail in \cite{book}.

\end{document}